\documentclass[12pt,preprint]{aastex}

\title{The Automated Palomar 60-Inch Telescope}
\author{S.~Bradley Cenko\altaffilmark{1}, Derek B.~Fox\altaffilmark{2},
	Dae-Sik Moon\altaffilmark{1,3}, Fiona A.~Harrison\altaffilmark{1},
	S.~R.~Kulkarni\altaffilmark{4}, John R.~Henning\altaffilmark{5},
	C.~Dani Guzman\altaffilmark{6}, Marco Bonati\altaffilmark{5},
	Roger M.~Smith\altaffilmark{5}, Robert P.~Thicksten\altaffilmark{5}, 
        Michael W.~Doyle\altaffilmark{5}, Hal L.~Petrie\altaffilmark{5},	
	Avishay Gal-Yam\altaffilmark{4,7}, Alicia M.~Soderberg\altaffilmark{4},
	Nathaniel L.~Anagnostou\altaffilmark{8},
	Anastasia C.~Laity\altaffilmark{8}}
\altaffiltext{1}{Space Radiation Laboratory, MS 220-47, 
	California Institute of Technology, Pasadena, CA 91125}
\altaffiltext{2}{Department of Astronomy \& Astrophysics, 525 Davey Laboratory,
	Pennsylvania State University, University Park, PA 16802}
\altaffiltext{3}{Robert A. Millikan Fellow}
\altaffiltext{4}{Department of Astronomy, MS 105-24, California Institute of
	Technology, Pasadena, CA 91125}
\altaffiltext{5}{Caltech Optical Observatories, MS 105-24, California 
	Institute of Technology, Pasadena, CA 91125}
\altaffiltext{6}{Gemini Observatory, c/o AURA, Casilla 603, La Serena, Chile}
\altaffiltext{7}{Hubble Fellow}
\altaffiltext{8}{Infrared Processing and Analysis Center, MC 100-22, 
	770 S.~Wilson Ave., Pasadena, CA 91125} 

\email{cenko@srl.caltech.edu}

\shorttitle{Automated Palomar 60-Inch Telescope}
\shortauthors{Cenko et al.}

\begin{document}

\newcommand{\Swift}{\textit{Swift}}
\newcommand{\Hete}{\textit{Hete-2}}
\newcommand{\Integral}{\textit{Integral}}
\newcommand{\Bepposax}{\textit{BeppoSAX}}

\begin{abstract} 
We have converted the Palomar 60-inch telescope (P60) from a classical
night assistant-operated telescope to a fully robotic facility. 
The automated system, which has been operational
since September 2004,  is designed for moderately fast ($t \lesssim 3$ minutes)
and sustained ($R \lesssim 23$ mag) observations of 
gamma-ray burst afterglows and other transient events.  Routine 
queue-scheduled observations can be interrupted in response
to electronic notification of transient events.   An automated pipeline
reduces data in real-time, which is then stored on a 
searchable web-based archive for ease of distribution.  We
describe here the design requirements, hardware and software upgrades,
and lessons learned from roboticization.  We present an overview of the
current system performance as well as plans for future upgrades.
\end{abstract}



\section{Introduction}
\label{sec:intro}
The field of optical transient astronomy has matured to produce 
numerous important scientific discoveries in recent years.
Type Ia supernovae (SNe) have been used as standard candles to produce
Hubble diagrams out to $z \sim 0.5$, providing evidence that the
expansion of the universe is accelerating \citep{rfc+98,pag+99}.
Observations of the broadband afterglows of long-duration ($t > 2$ s)
gamma-ray bursts (GRBs) have revealed an association with the deaths
of super-massive stars \citep{gvp+98,smg+03,hsm+03}.  The discovery of the 
first afterglows and host galaxies of short-duration ($t < 2$ s) GRBs 
\citep{gso+05,bpp+06,hwf+05,ffp+05}
has possibly revealed a new class of GRB progenitors: compact binary
coalescence \citep{elp+89}.

As interest in the field has steadily grown, new, more powerful methods of
identifying optical transients have been developed.  
The \Swift\ Gamma-ray Burst Explorer \citep{gcg+04}
is currently providing $\sim$ 100 prompt GRB
localizations per year, an order-of-magnitude improvement over previous
missions.  Planned wide-angle, high-cadence surveys with large facilities,
such as Pan-STARRS \citep{kab+02} and LSST \citep{t05}, promise to overwhelm
our current follow-up capability, providing  
hundreds of variable optical sources each night.  

Dedicated, robotic, medium aperture ($1 - 3$ m) telescopes have
the opportunity over the next few years to play
a crucial role in this field.  Like small-aperture ($< 0.5$-m), robotic
facilities, they can respond autonomously to transient alerts, providing
observations at early times.  And given the relative abundance of such 
telescopes, it is entirely feasible to focus predominantly on transient
astronomy.  However, like larger telescope ($> 5$-m), interesting
events can be followed for longer durations and in multiple colors.
In this sense robotic, medium-aperture facilities can act to bridge the gap 
between the earliest rapid-response observations and deep, late-time imaging
and spectroscopy.

\begin{figure}[p]
     \plotone{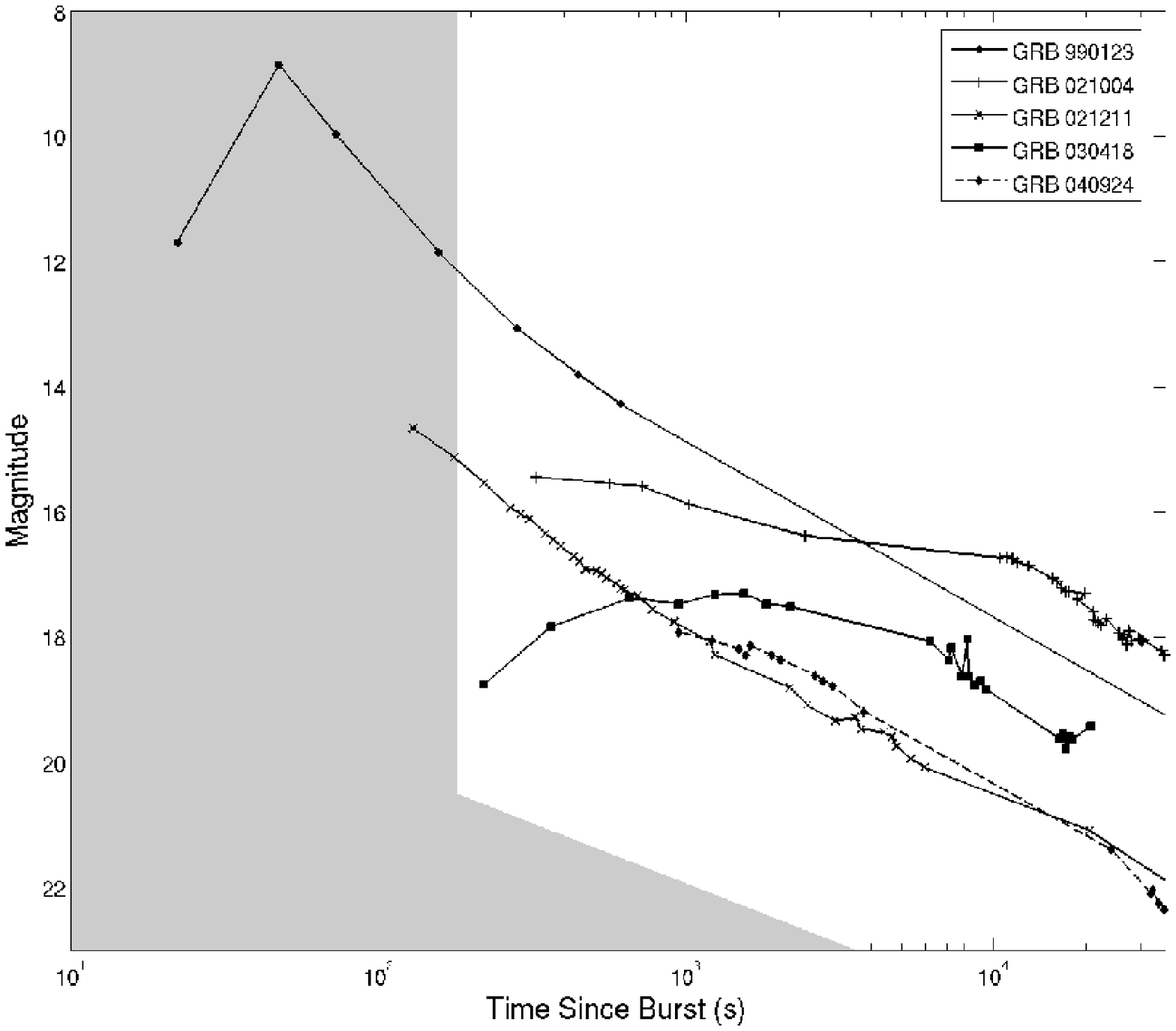}
     \caption[Early Afterglows of pre-\Swift\
     GRBs and P60 Response Capabilities]
     {Early Afterglows of pre-\Swift\
     GRBs and P60 Response Capabilities.
     Regions with a white background are accessible for automated P60
     observations: $t \gtrsim 3$ minutes, $R \lesssim 23$ mag.
     With only
     a handful of examples, the early optical afterglows of pre-\Swift\ GRBs
     show a marked diversity.  GRB\,990123 \citep{abb+99} and
     GRB\,021211 \citep{fps+03,lfc+03} exhibit the fast $t^{-2}$ early-time
     decay indicative of adiabatic evolution of the reverse shock.
     On the other hand, GRB\,021004
     \citep{fyk+03,hwf+03,psr+03} shows a distinctive slow $t^{-0.4}$ decay
     that likely signifies continuing energy input to shock regions.
     Reverse shock emission from GRB\,030418 \citep{rsp+04} was not seen;
     the optical peak at $t=0.4$ hours is due to the forward shock
     component.
     As a proof of concept, the P60 was the first to report the
     afterglow of GRB\,040924 \citep{fm04,lfc+04,hlh+04,sab+04,
     kay+04}.  The
     early time behavior is quite similar to that of GRB\,021211.}
\label{fig:earlylc}
\end{figure}

\begin{figure}[p]
     \plotone{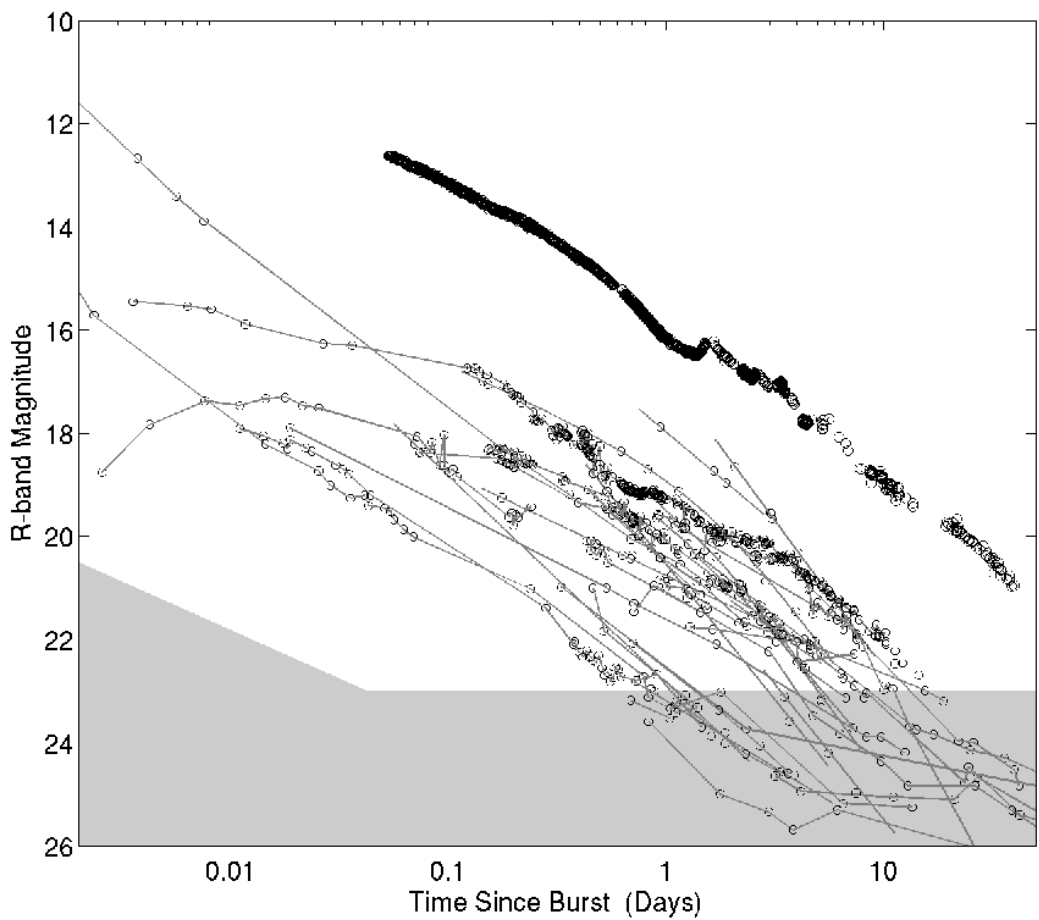}
     \caption[Late-time light curves of pre-\Swift\ GRB afterglows]
     {Late-time light
     curves of pre-\Swift\ GRB afterglows.
     The gray shaded region displays the phase space inaccessible to automated
     P60 observations.  Observations of most afterglows require
     $> 1$-m class facilities after the first night; investigation of
     optically-extinguished ("dark") or high-redshift bursts require
     such facilities merely to register detections or collect physically
     interesting upper limits.}
\label{fig:latelc}
\end{figure}

To this end, we have roboticized the Palomar 60-inch telescope (P60).  
As a dedicated, robotic facility, the P60 is capable of responding 
moderately fast ($t \lesssim 3$ min) to transient alerts.  With
the increased event rate of \Swift, the P60 is providing observations of 
the poorly understood early afterglow phase (Fig. \ref{fig:earlylc}).
Additionally, as a 1.5-m telescope, the P60 can continue the sequence of 
observations longer than most robotic telescopes.  
As Figure \ref{fig:latelc} shows, 
one day after the burst, most afterglows have faded below $R = 20$; however,
for days or even weeks after that, they remain at levels of $R < 23$ 
accessible to P60 photometry.

In this work, we first outline the high-level design
requirements of a robotic system optimized for observations of
transient sources (\S\ref{sec:design}).
\S\ref{sec:procedure} provides the details of the automation procedure,
including both the hardware and the software efforts. \S 
\ref{sec:performance} describes the current system performance (as of
May 2006), which will primarily be of use for those interested in 
observing with the P60.  
Finally, in \S\ref{sec:conclusions}, we conclude with a summary of the project
status and a discussion of possible future improvements to the robotic
system.

\section{General Design Considerations} 
\label{sec:design} 
Designing a robotic system for transient astronomy presents 
a unique set of challenges
from both a hardware and a software perspective.  It is
necessary to create an intelligent system that can reliably handle the
roles usually provided by the observer and night assistant at a standard
facility (see, e.g., \citealt{gh89}).

\begin{figure}[p]
     \plotone{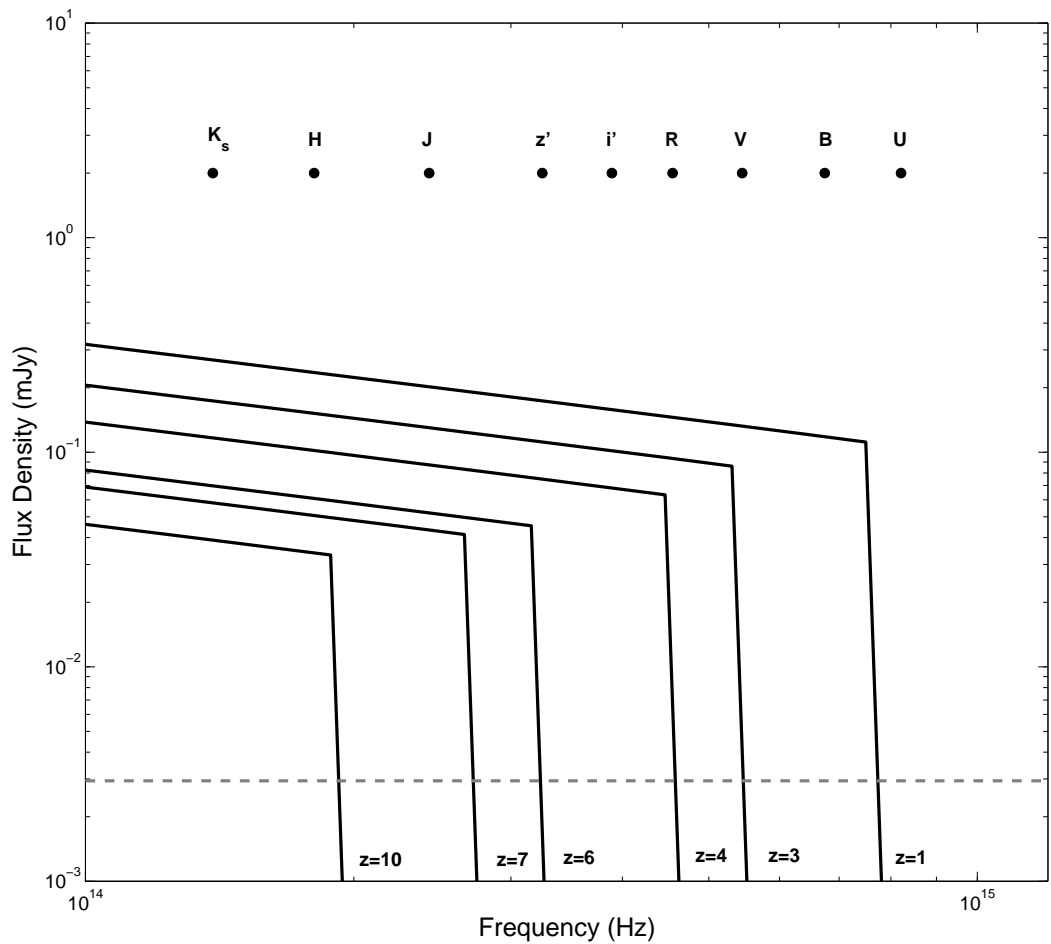}
     \caption{Optical and Near-Infrared SEDs of GRB
     Afterglows as a Function of Redshift.  These SEDs are models of the
     afterglow of GRB\,990510
     one hour after the burst \citep{pk01},
     viewed at redshifts ranging
     from $z = 1-10$.  The P60 $R$-band sensitivity (one hour integration,
     $R \approx 23$ mag) is shown as a dashed line, extended to all frequencies
     for reference.  The central wavelengths of the
     broadband filters on the P60 are drawn above the spectra, as well
     as the standard $J\,H\,K_{\mathrm{s}}$ near-infrared filter set.
     Lyman-$\alpha$ absorption in the
     IGM causes the steep cut-off in the afterglow spectra, which can
     be used to estimate the redshift of GRB afterglows photometrically
     \citep{lr00}.}
\label{fig:highz}
\end{figure}

Given our scientific objectives, we identified following system
requirements for the Palomar 60-inch Automation Project:

\begin{enumerate}
\item \textbf{Automated transient response in $\lesssim 3$ min}.  
GRB afterglows are predicted
to decay in time as a power-law ($F_{\nu} \propto t^{-\alpha}$) with index
$\alpha \approx$ 1--2, depending on whether the emission is dominated
by the forward ($\alpha_{\mathrm{FS}} \approx 1$; \citealt{spn98}) or 
reverse ($\alpha_{\mathrm{RS}} \approx 2$; \citealt{sp99}) shock.  
For (optically) bright bursts, rapid response enables studies of
the afterglow at its brightest, shedding light on the poorly
understood early afterglow phase (Figure~\ref{fig:earlylc}).   
For the fainter bursts, rapid response is 
required simply to obtain a detection or even a meaningful upper limit (Figure
\ref{fig:latelc}).  Our desired response overhead is limited 
primarily by the telescope slew time.

\item \textbf{CCD Readout in $< 30$ s}.  
Given the expected power-law behavior, densely-sampled
observations are necessary to accurately characterize 
the early afterglow decay.  And since our current system is not equipped with
an automated guider, deep observations must be broken down into many
individual exposures (and hence many accompanying readouts).  
Given typical values for our telescope slew 
(3 min) and exposure (1--3 min) times, we determined a readout time $< 30$ s
would not significantly affect our sampling rate or efficiency.  

\item \textbf{Photometry from the near ultra-violet to the 
near-infrared}:  GRB redshifts can be estimated photometrically by 
modeling afterglow spectral energy distributions (SEDs).  Ly-$\alpha$
absorption in the Inter-Galactic Medium (IGM) causes a steep cut-off in 
the SED, the location of which indicates the afterglow redshift \citep{lr00}.  
To constrain as large a spectral range as 
possible ($2 < z < 6$), we require coverage over the entire optical bandpass
(see Figure~\ref{fig:highz}).   
The ideal solution would be a multi-band camera,
providing simultaneous imaging in multiple filters.  
The cost of either purchasing or building such an instrument, however, was
too high for our first generation of operations.
Instead, we employ a 12-position filter wheel, 
with coverage spanning from Johnson $U$-band ($\lambda_{\mathrm{c}} = 3652$
\AA) to Sloan $z^{\prime}$ ($\lambda_{\mathrm{c}} = 9222$ \AA). 

\item \textbf{Intelligent observation oversight}.  Like a 
virtual night assistant, a centralized source of information
is required to effectively manage nightly observations (i.e.~telescope, 
weather, and instrument status information).  Under ideal conditions, this is
not a difficult task.  More challenging, however, is implementing
a robust capability to intelligently respond
to adverse conditions.  

\item \textbf{Queue scheduling system for non-TOO mode}.
Since not all of the telescope time is devoted to rapid-response GRB
observations, a
scheduler is needed to handle standard scientific observations, as well as 
calibration images.  We chose to implement a queue-scheduler,
as it is capable of providing 
real-time management of observations (i.e.~targets can 
be submitted to the queue at any time) with a minimal amount
of daily oversight (night-to-night memory ensures that there is no need
to write daily target lists).  Furthermore, a queue-scheduler is ideally suited
for long-term monitoring of transient objects; SNe and GRBs can be left in
the queue for regular monitoring on time scales of weeks or even months.  

\item \textbf{Automated, real-time ($< 2$ min) data reduction}.  
Real-time data reduction
is necessary for several reasons.  First and foremost, feedback is required
for standard system oversight commonly performed by observers present
at the telescope.  Focusing is the simplest example.
Secondly, rapid identification of 
optical counterparts is critical for intelligent follow-up observations.
High-resolution absorption spectroscopy in particular requires a rapid
turn-around with large facilities. 
Finally, properly handling the large amounts of data produced on a nightly
basis requires that data reduction be fully automated. 

\item \textbf{Fully searchable, web-based data archive}.  
The average P60 data rate, including daily calibration files,
 is $\sim$ 5 GB per night.  Furthermore, with our queue-scheduling system,
science images are obtained for a large number of users ($\sim 10$) on most 
nights.  We therefore opted for a high-capacity, fully searchable data archive
for ease of data storage and distribution.

\end{enumerate}

\section{Automation Procedure}
\label{sec:procedure}
In the previous section we outlined the design 
requirements for the automated system.
Here we describe the techniques we 
have used to meet these requirements in a more thorough manner.


\subsection{New CCD \& Electronics}
\label{sec:newccd}

The previous P60 CCD took almost three minutes to read out, unacceptably
long given our desired response time of $\lesssim 3$ minutes.  Furthermore,
the camera was only accessible via a local microVAX terminal, making
automated observations impossible.  To meet our design requirements,
we chose to build a new camera using the latest San Diego State
University controller, Generation III electronics (SDSU-III; \citealt{ll00}).
This system is capable of
better performance than an off-the-shelf product, with the trade-off that
a significant time investment was required for development and testing.
In the following two sections, we describe the new electronics 
(\S\ref{sec:electronics}) and the software used to control the camera
(ArcVIEW; \S\ref{sec:arcview}).

\subsubsection{SDSU-III Electronics}
\label{sec:electronics}
\begin{figure}[p]
     \plotone{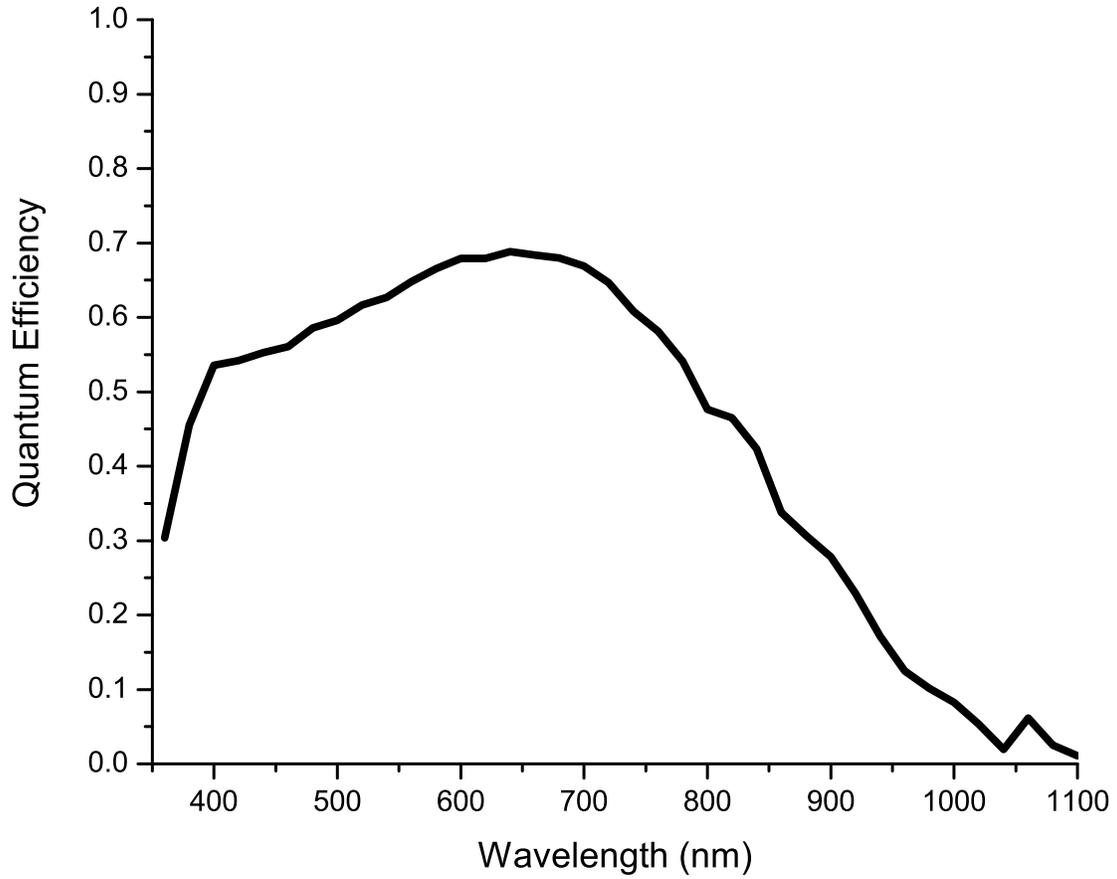}
     \caption[Old P60 CCD Quantum Efficiency]
     {Previous P60 CCD Quantum Efficiency.  While we have not measured the
     quantum efficiency of the new P60 CCD, it is identical in design to the
     previous version shown here.  Comparing observations made with both
     detectors indicates a comparable overall performance.}
\label{fig:ccd13qe}
\end{figure}

The telescope was equipped with a new SITe 2K $\times$ 2K 
back-illuminated CCD.  While we have not measured the quantum efficiency
of the new device, our observations indicate its quantum efficiency
is comparable to that of the previous camera (which was an identical
SITe 2k $\times$ 2k CCD).  For reference, we include a quantum efficiency
plot from the old CCD in Figure \ref{fig:ccd13qe}.

The new CCD is controlled by an SDSU-III controller \citep{ll00}.
The new controller contains a 
faster optical link than the Generation II system, as well as a 
newly designed timing board.  The system is capable of reading out four 
channels in parallel.  However, to reduce costs and simplify fabrication,
we currently utilize only two amplifiers for readout.

Temperature sensors were placed in thermal contact with the CCD, 
the dewar neck and 
can, as well as on board the electronics.  These sensors are capable of
triggering an alarm under abnormal conditions, for example when the
dewar runs out of liquid nitrogen and begins to warm. 

In addition to the standard full-frame readout mode, two additional 
capabilities have been implemented.  Using the region-of-interest (ROI) 
functionality, 
we can read out only a subsection of the chip.  This is particularly
important for small GRB error circles, helping to improve both the sampling
rate and efficiency of our system. 
Additionally, the ability to manipulate charge independent of the readout
(``parallel shift'') greatly decreases the time required for a focus loop.
This has been of utmost importance, given the difficulties we have 
encountered maintaining system focus throughout the night (see 
\S\ref{sec:efficiency}). 

The relevant characteristics of the new camera are outlined in Tables
\ref{tab:detector} and \ref{tab:readout}.   
The P60 camera was the first developed under an engineering scheme 
designed to standardize enclosures and cabling for new instruments on 
the mountain.  The lessons learned have been extended to future 
instruments being developed for Palomar Observatory.

\begin{deluxetable}{lrrr}
  \tabletypesize{\footnotesize}
  \tablecaption{New P60 CCD \& Electronics Capabilities.}
  \tablecolumns{4}
  \tablewidth{0pc}
  \tablehead{\colhead{Property} & \colhead{Amplifier 1} &
            \colhead{Amplifier 2} & \colhead{Full Chip}
            }
  \startdata
       Array Size & $2048 \times 1024$ & $2048 \times 1024$ & $2048 \times
                2048$ \\
       Pixel Size ($\mu$m) & \nodata & \nodata & 24 \\
       Plate Scale (arcsec pixel$^{-1}$) & \nodata & \nodata & 0.378 \\
       Field of View (arcmin) & $12.9 \times 6.5$ & $12.9 \times 6.5$
                & $12.9 \times 12.9$ \\
       Gain (e$^{-}$ ADU$^{-1}$) & 2.2 & 2.8 & \nodata \\
       Read Noise (e$^{-}$) & 5.3 & 7.8 & \nodata \\
       Dark Current (e$^{-}$ s$^{-1}$) & $10^{-3}$ & $10^{-3}$ & \nodata \\
       Charge Transfer Efficiency & $>$ 99.999\% & 99.999\% & \nodata \\
       Full Well Capacity (e$^{-}$) & 130,000 & 140,000 & \nodata \\
       Bias Level (ADU) & 610 & 445 & \nodata \\
       Saturation Limit (ADU) & 50,000 & 45,000 & \nodata \\
  \enddata
\label{tab:detector}
\end{deluxetable}

\begin{deluxetable}{lrr}
  \tabletypesize{\footnotesize}
  \tablecaption{P60 CCD Readout Time}
  \tablecolumns{3}
  \tablewidth{0pc}
  \tablehead{\colhead{Fraction of Array} & \colhead{Sky Size} &
             \colhead{Readout Time} \\ & \colhead{(arcmin)} & \colhead{(s)}
            }
  \startdata
        Full & $12.9 \times 12.9$ & 24 \\
        $\frac{1}{2}$ & $6.5 \times 12.9$ & 18 \\
        $\frac{1}{4}$ & $6.5 \times 6.5$ & 10 \\
  \enddata
\label{tab:readout}
\end{deluxetable}

\subsubsection{Instrument Control System: ArcVIEW}
\label{sec:arcview}
The software used to control instrument operation is called
ArcVIEW, a package that was developed at 
the Cerro Tololo Inter-American Observatory and Caltech. 
It is based on Labview
(interfaces and communication) and C (real time data processing and drivers
API).  

The ArcVIEW architecture consists of a set of software modules that can be
loaded or unloaded dynamically to control different processes. 
The core of the software receives commands and passes them to the 
appropriate module for processing.  
A translation layer built into the system allows
for transparent hardware control (i.e.~the standard command set available
to the user is independent of the details of the hardware
being controlled).  

ArcVIEW commands are sent as plain ASCII strings 
passed through raw sockets. 
Graphical User Interfaces (GUIs) are not needed to control the system;
however, some of them  are provided in order to handle data taking, filter
movements, TCS commands, and  low-level engineering commands in a
user-friendly way.

Besides the normal command/response channel, ArcVIEW contains
an optional asynchronous
message channel, that allows the system to send asynchronous alarm
messages (temperatures, power supplies, etc.), callbacks, or event messages
to the connected client.  Using this extra channel it is possible to perform 
simultaneous actions (e.g.~moving the telescope while reading out the array).

The final output of the system is an image (or
sequence of images) written in FITS format and containing user-defined
header information.  The two P60 amplifiers are read out
and stored as a multi-extension FITS file.  

\begin{figure}[p]
     \plotone{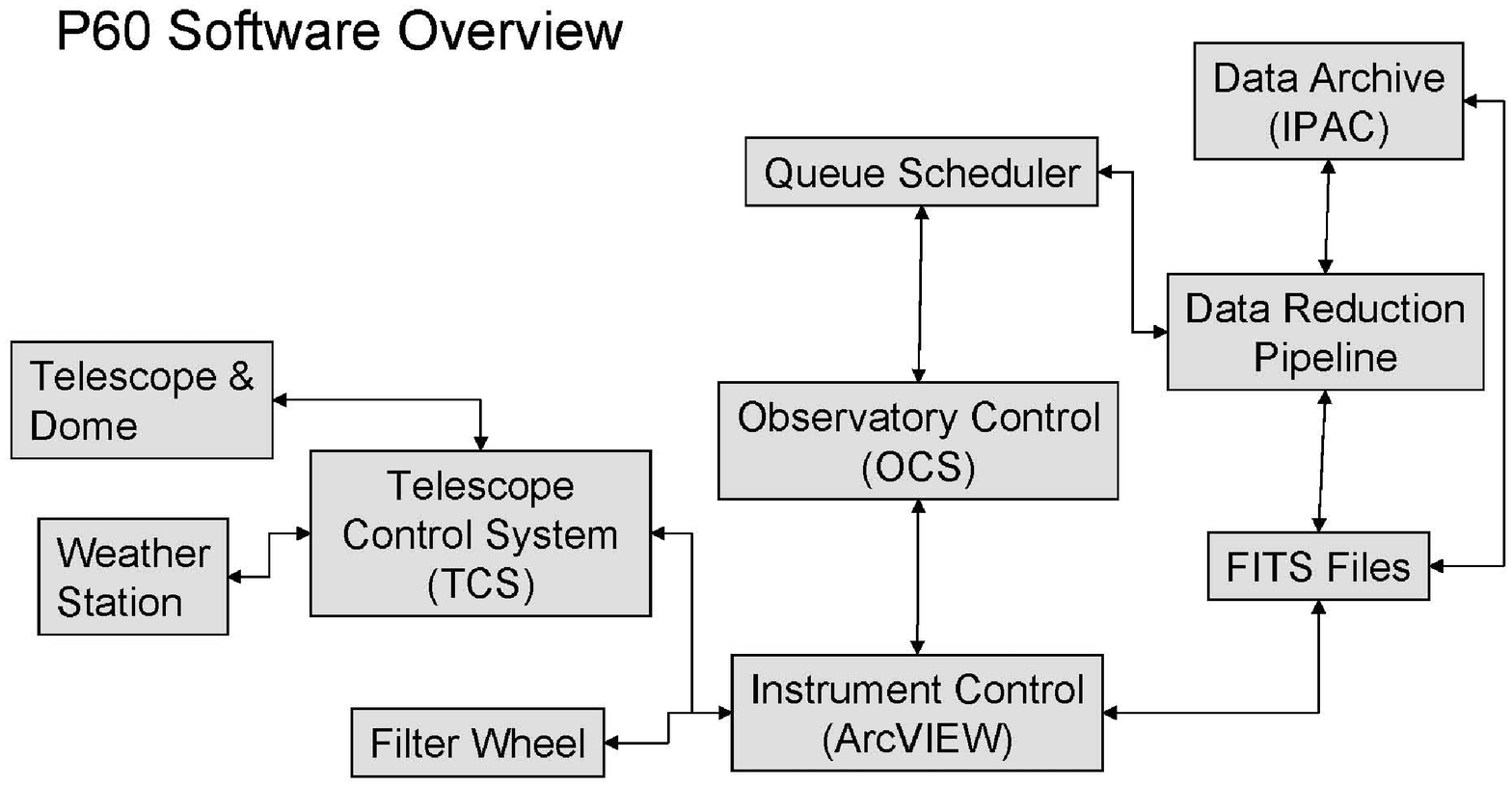}
     \caption[P60 Software Overview]{P60 Software Overview.  Arrows indicate
     direct channels of communication.  The modular design was chosen to ensure
     both stability and ease of upgrade / repair.}
\label{fig:software}
\end{figure}

We have chosen a modular design for our major software components, as
illustrated in Figure \ref{fig:software}.  Each component acts
independently, with a well-defined communication protocol between the different
modules.  This makes software upgrades easier, allows for a clean division
of labor and responsibilities, and guarantees a more robust system, as 
failure in one component does not necessarily imply complete system failure.
Modular designs have long been in use at automated facilities and have proved
both reliable and effective (see, e.g., \citealt{ht92,sc97,gws01,bsb+05}).

On the P60, ArcVIEW acts as a single point of contact between hardware 
operation (telescope, CCD, and filter wheel), and all other system components
(see Figure \ref{fig:software}).  

\subsection{Observatory Control System}
\label{sec:OCS}
The purpose of the Observatory Control System (OCS) is to provide intelligent 
oversight of nightly observations and to coordinate information from 
all system components (Fig.~\ref{fig:software}).  
We identify four primary tasks for which the OCS is responsible, each discussed
below.

First, at the beginning of each night, the OCS spawns the queue-scheduling
software in a separate process (see \S\ref{sec:OSS}).  These two systems
communicate throughout the night via a socket, as real-time target selection 
depends on the success of previous observations.  

After receiving an observation request, the OCS is then responsible for
executing it in a safe and efficient manner.  Communication with the Telescope
Control System (TCS), via the transparent ArcVIEW intermediary, ensures
that external conditions permit the requested observation.  All component
tasks that can be completed in parallel (e.g.~moving the telescope and
filter wheel) are done so to improve system efficiency.  An observation
is considered to have completed successfully when the readout of the final
exposure begins.

Third, after the successful completion of the first images on any given night,
the OCS spawns the data reduction pipeline in a separate process (see
\S\ref{sec:reduction}).  These two systems communicate to ensure the integrity
of science images, most notably by maintaining telescope focus
throughout the night (see \S\ref{sec:efficiency}).  

Finally, the OCS handles
any errors that arise during the normal course
of operations.  Each error condition is assigned a level in a hierarchy
of functionality.  Lower levels correspond to more basic, elementary 
functionality, and vice versa.  When an error is discovered, the OCS will begin
at the appropriate error level and work downward until the depth of the
error condition is determined.  The OCS then works to restore the system to 
functionality.  If no solution can be found, the system goes into a safe mode,
closing the dome and terminating observations.  Email notices and text 
messages are sent to alert users of this condition.  

As an example, we consider an error generated by the focus encoder during
routine operation.  The OCS first verifies communication with the TCS.
If this fails and cannot be restored, the system checks communication with 
ArcVIEW, as it is responsible for routing most communication.  If this too
fails and cannot be restarted, the OCS checks for internet connectivity.  
This process continues until either a solution is discovered or human 
intervention is required.  Similar systems have been utilized successfully
on other automated facilities \citep{ht92,gws01}.

\subsection{Observation Scheduling System}
\label{sec:OSS}

In the design of the Observation Schedule System (OSS), 
we have deliberately pursued a
``short-sighted'' strategy of selecting targets in real-time. 
That is, observations are chosen at each point in the night when the OCS
reports being in a ready state -- rather than attempting to optimize a
sequence of observations over the course of a full night (or over
multiple nights).  This strategy is relatively well suited to
ground-based observations, where future observing conditions are
unknown and observing overheads are a relatively minor concern.
Moreover, the scheduling protocol and target list for P60 observations
are modest enough that a full evaluation of the target list can be
performed in a matter of seconds.  This principle of ``just in time''
scheduling has also been pursued at several larger-scale
queue-observing facilities \citep{cgs+98,skk+00,atb+04}, as well as more
modest robotic observatories \citep{hvw+90,fs04}.

Target scores are determined on the basis of raw target
priorities, which are fixed in advance, combined with the application
of several parametric weightings.  The most important of these for
scheduling purposes are the \textit{Airmass} and \textit{Night} 
weighting variables,
which take as input the current airmass of the target and the number
of hours left before the target becomes unobservable (due to
target-set or morning twilight), respectively.

The nature of the effect of each weighting is the same.  Based on the
value of the input variable, the weight is calculated and applied as a
multiplier to the target score (initially, the target priority).  If
the weighting is found to be zero then the target score is necessarily
zero; otherwise, the target score will be increased or decreased
depending on whether the weight in question is calculated to be
greater or less than one.

The full list of possible weighting variables includes:
\begin{itemize}
  \item \textit{Airmass}, with input variable the current airmass of the
        target.  This weighting prefers sources that are close to
        transit (minimum airmass).
  \item \textit{Night}, with input variable the number of hours until the
        source becomes unobservable.  This weighting helps ensure
        efficiency of the scheduler operations since it prefers
        sources that are setting rather than rising.  The estimated
        duration of the target's full exposure sequence is included in
        the calculation.
  \item \textit{Moondeg}, with input variable 180 degrees minus the current
        angular distance from the target to the moon.  This avoids
        taking images with high sky background due to moonlight.
  \item \textit{Seeing}, with input variable the current seeing in arcseconds.
        This allows the segregation of programs according to whether
        their science is adversely affected by poor seeing.
  \item \textit{Extinction}, with input variable the current magnitude of
        extinction, in the $R$ band, due to clouds.  This
        allows segregation of programs according to how strongly they
        are affected by reduced sensitivity.
\end{itemize}
The \textit{Seeing} and \textit{Extinction} 
weightings are not yet in operation,
but should be applied dynamically within the OSS by the end of summer 
2006.

In addition to these parametric weightings, target scores are also
adjusted based on timing criteria.  The default logarithmic timing scheme
steadily increases the score of a target from night to night until it
has been observed.  Alternate timing schemes allow for periodic
(ephemeris-based) or regular aperiodic (``best effort'') monitoring of
targets, or for target activation within a specified window of time
only.

Finally, we have found it important to increase the score of targets
once they have been observed on a given night, so that they are more
likely to be observed to completion (one or more sets of the requested
exposure sequence) during that night.  This prevents fragmentation of
observer programs, and reduces overheads which are mostly incurred on
a per-target basis.

\begin{figure}[p]
   \plottwo{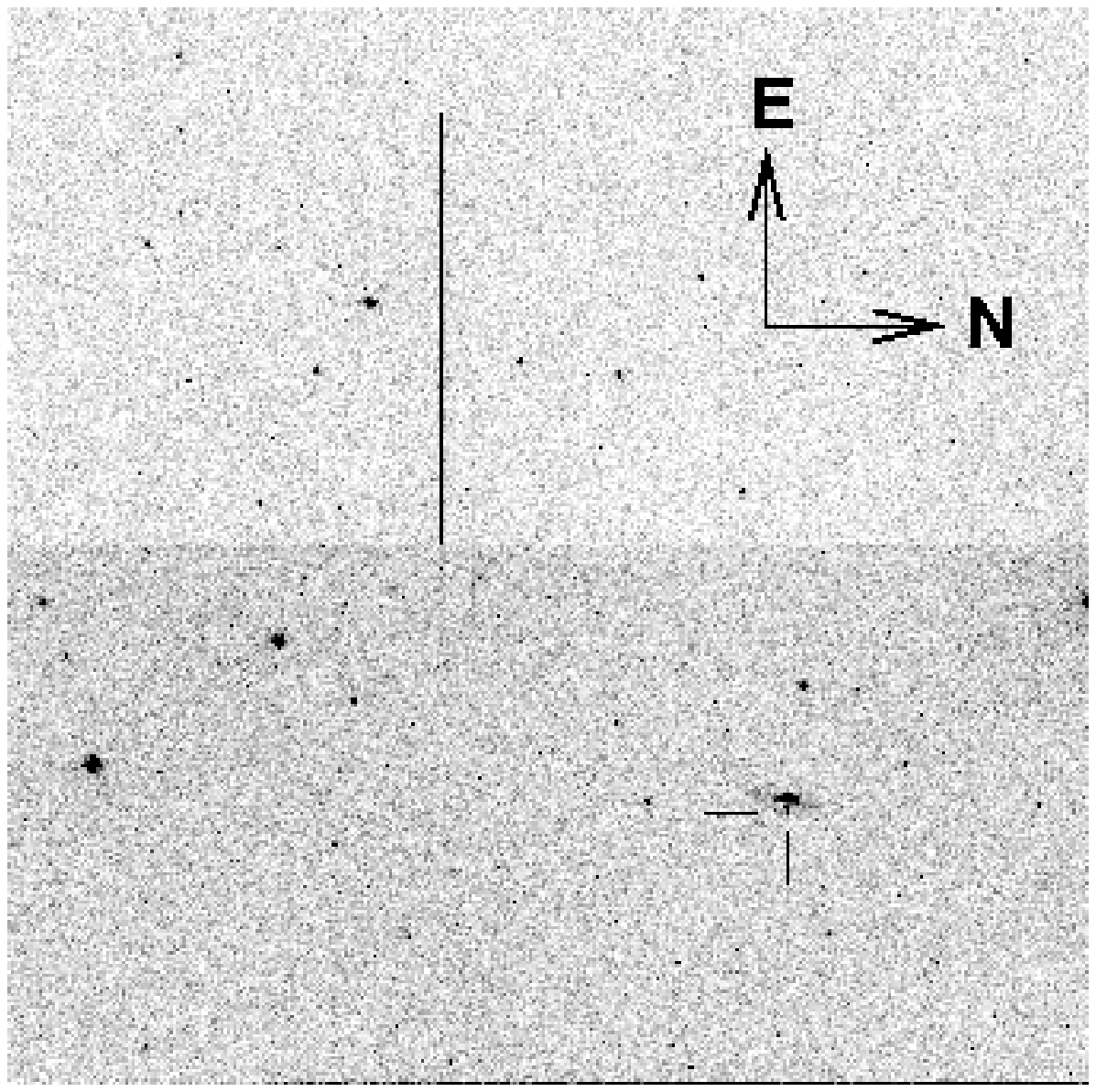}{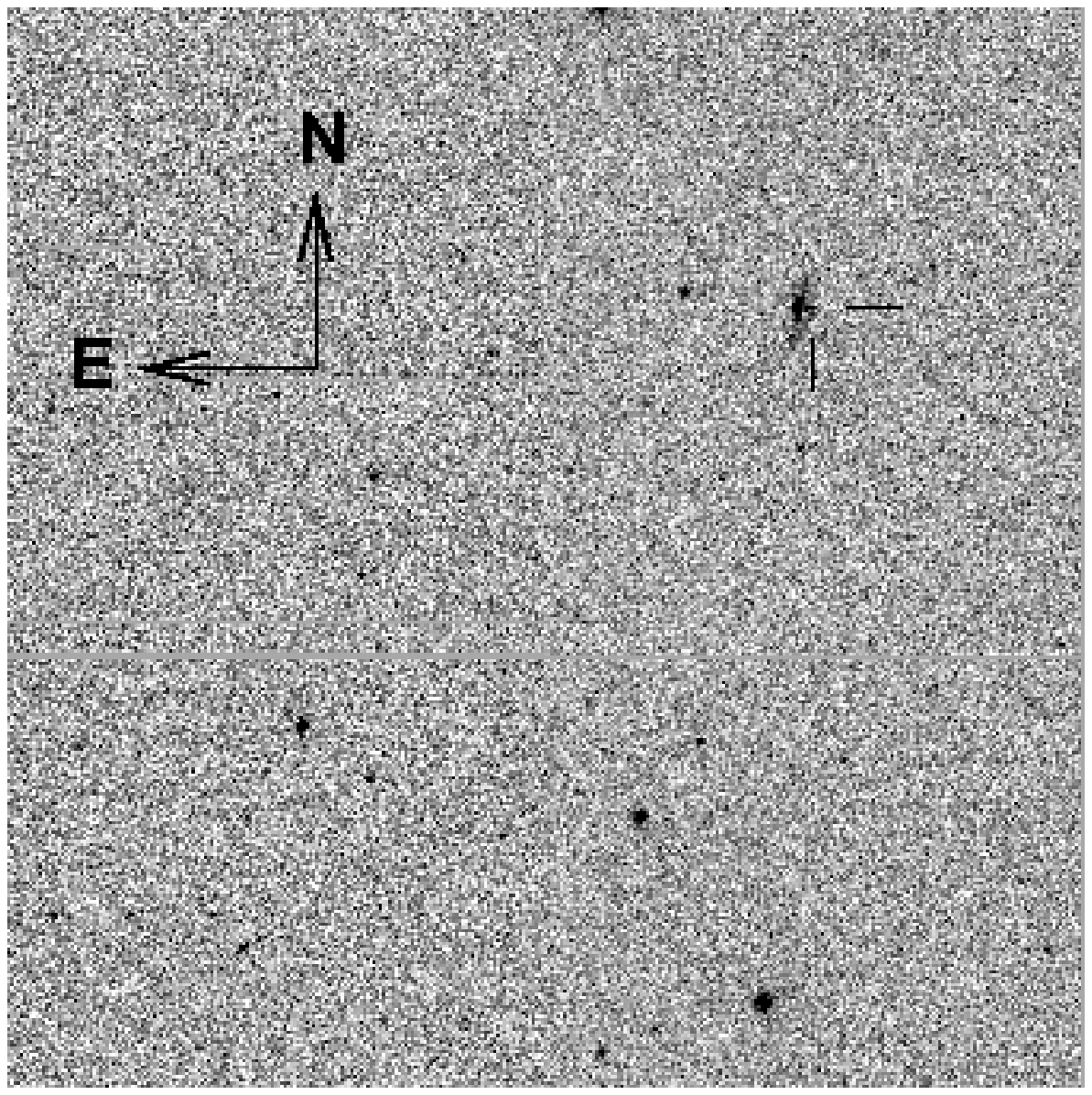}
   \caption{P60 On-Sky Images.  \textit{Left}:  Raw P60 image of SN2006be.
   The object, located just West of its host galaxy IC 4582, is indicated
   with the two black tick marks.  The row of bad columns is clearly
   visible on the top amplifier.  Because of these cosmetic defects and the
   higher read noise of the top amplifier, we recommend a small coordinate
   offset (3\arcmin\ N, 3\arcmin\ W) for non-extended sources, as has been
   applied for this object.  \textit{Right}:  Processed P60 image of
   SN 2006be.  Here we display the
   output of the real-time data reduction pipeline, as
   described in \S\ref{sec:reduction}.
   The image has been rotated to the standard
   orientation of North up and East to the left.}
\label{fig:p60ccd}
\end{figure}

\subsection{Image Analysis Pipeline}
\label{sec:reduction}
The constituent routines for our image analysis pipeline are composed
within the context of Pyraf\footnote{http://pyraf.stsci.edu/}, a
Python wrapper for the IRAF data reduction environment of the
NOAO\footnote{http://iraf.noao.edu/}.  The pipeline is instantiated in
a single Python script which can be run from the Linux command line.
The script runs continuously throughout the night, identifying new raw
images as they are copied into the target directory, and processing
them in real-time.

Pyraf allows access to IRAF routines from within Python, a scriptable,
object-oriented, high-level language environment.  In particular,
Python performs active memory management and, with its various
included modules, supports mathematical and logical operations on
array variables, regular-expression matching against text strings, and
easy access to FITS headers and data.  

Python scripts which access
arbitrary Pyraf routines can be executed from the command line.  The
speed of these scripts is not as fast as compiled C routines.
However, the single most substantial overhead for script execution is
incurred at startup as the Pyraf libraries (including IRAF) are loaded
into memory.  Once cached in memory, the speed of execution of our
scripts is competitive with native IRAF and adequate to our purposes.

The routines of the P60 pipeline execute the following reduction steps
in sequence: (1) De-mosaicking, which performs overscan-subtraction on
the separate image extensions produced by the two amplifiers, and
combines them into a monolithic image while preserving the values of
unique header keywords associated with each extension; (2) Bias
subtraction against our nightly bias image; (3) Flat-fielding against
the dome-flat images taken during the afternoon or previous morning,
sky-subtraction, and addition of the dead-reckoning world-coordinate
system (WCS); (4) Masking
of bad pixels, using the nightly bad pixel mask; (5) Object detection,
using a spawned
Sextractor\footnote{http://terapix.iap.fr/soft/sextractor/} process;
(6) WCS refinement via triangle-matching against the USNO B-1.0
catalog\footnote{http://www.nofs.navy.mil/data/FchPix/}, using 
the ASCFIT software \citep{jrb+02}; (7) Seeing and zero-point
estimation using USNO B-1.0 catalog stars identified in the image.

If an insufficient number
of stars are identified during the WCS refinement process for an
image, then the dead-reckoning WCS is left untouched and the seeing
and zero-point estimation steps are skipped.
Calibration products are produced from raw calibration bias and
dome-flat images at the start of the night as a separate process.

The final analysis task, which is performed by a special
single-purpose script, is to determine our best focus value and
current seeing from a single focus-run (multiple exposures and a
single readout) on a bright star.  For the sake of speed, this task
omits most of the standard processing steps.

Additional routines have been coded but are not run in an
automated fashion, either because of difficulty in robustly defining
their operations, or because of excessive processing requirements.  These
include: fringe image creation and defringing of $I$ and $z^{\prime}$-band
images; co-addition of multiple dithered images to achieve greater
depth of field; and mosaic co-addition of multiple images, using
Swarp\footnote{http://terapix.iap.fr/soft/swarp/}, to cover areas
significantly larger than the CCD field of view.

The P60 pipeline routines are general and can be readily applied to
other data reduction tasks; indeed, we have already adapted them to
the construction of an interactive pipeline for Wide-Field
Infrared Camera (WIRC; \citealt{weh+03}) data reduction at
the Hale 200'' Telescope.

\subsection{Data Archive}
\label{sec:archive}
The P60 data archive is designed to securely store data collected at the
robotic facility, and to provide efficient and convenient access to users from
the P60 partner institutions.  
In return for a $10\%$ share of telescope time, 
the Infrared Processing and Analysis Center (IPAC)
has assumed responsibility for the
procurement, installation and maintenance of the archive hardware, as well as
for database software development, following specifications provided by the
P60 science team at Caltech.

The archive routinely stores the entire set of raw frames, calibration data,
and pipeline-processed images collected nightly at the telescope. The data are
transmitted down from Palomar mountain to the Caltech campus over the new
HPWREN fast data link. The images are transmitted in a non-lossy compressed
form, and MD5 checksums are used to verify their integrity. At IPAC, all files
are stored on a cluster of Sun computers hosting the archive server and
database structure. A RAID5 NEXSAN Ataboy disk farm provides approximately 3
TB of disk space.  A second copy of the data is kept on 
Caltech computers at Robinson Lab as backup. Each nightly batch of data is
ingested into the database software, which has an astronomy-optimized
architecture similar to other IRSA archives. User access is provided through a
web-based interface.  
Using the archive webpage, users can query the database, locate data they
require, and request it from the archive. Data delivery is from a staging
area, following email notification to the user. Under normal operating
conditions, small data packets can be obtained in this way within minutes.

\section{Automated System Performance}
\label{sec:performance}
The P60 has been running in a fully-automated mode since September 2004.  
This includes all aspects of operation, from the automated queue-scheduler
through nightly ingest of archival data.  Here we present 
an overview of the
current system performance, focusing primarily on information relevant
for interested P60 observers. 

\subsection{CCD Camera, Telescope, and Filters}
\label{sec:ccdperformance}
At the current date (June 2006), the camera is performing
reliably and  
has met all relevant specifications.  
Since the fall of 2004, the amount of time lost due to
detector or electronics problems (or related software) is small 
($< 5$\%).  A summary of the relevant camera details can be found  
in Tables \ref{tab:detector} and \ref{tab:readout}.

The most relevant characteristic for our science goals is the readout time. 
The full frame readout time of the system is 24 seconds.  
This can be significantly reduced, however, by using the region-of-interest
mode (\S\ref{sec:electronics}).  For instance,
a 6\arcmin\ by 6\arcmin\ field ($\frac{1}{4}$ of the chip) requires 
only 10 seconds to read out.  

We have found amplifier 1 (the ``bottom'' amplifier) 
has a significantly lower read noise than amplifier 2 (the ``top'' amplifier,
5.3 vs.~7.8 e$^{-}$). The top region of the CCD
is also cosmetically less pleasing
than the bottom region, as several adjacent bright columns run through
the center portion of the CCD (see Figure \ref{fig:p60ccd}).  
We therefore recommend applying
a small offset from the central location ($+3'$ RA, $-3'$ Dec) for 
non-extended sources.  We have added an optional offset parameter to our 
target specification protocol to make this change easier for users. 

The pointing accuracy of the system is more than sufficient for our needs,
with typical RMS values of 15\arcsec.  We have found, however, somewhat deviant
behavior (up to 45\arcsec\ offsets) for targets observed at large airmass
($> 3$).  We believe this is caused by different pointing behavior with the
eyepiece mounted (used for rapid manual calculation of the pointing model) 
than with the CCD camera mounted (nightly observations).  We are currently 
investigating this issue in more depth.  However, we note that given our 
large field-of-view, even pointing errors as large as 1\arcmin\ are unlikely 
to cause significant problems.

\begin{figure}[tb]
     \plottwo{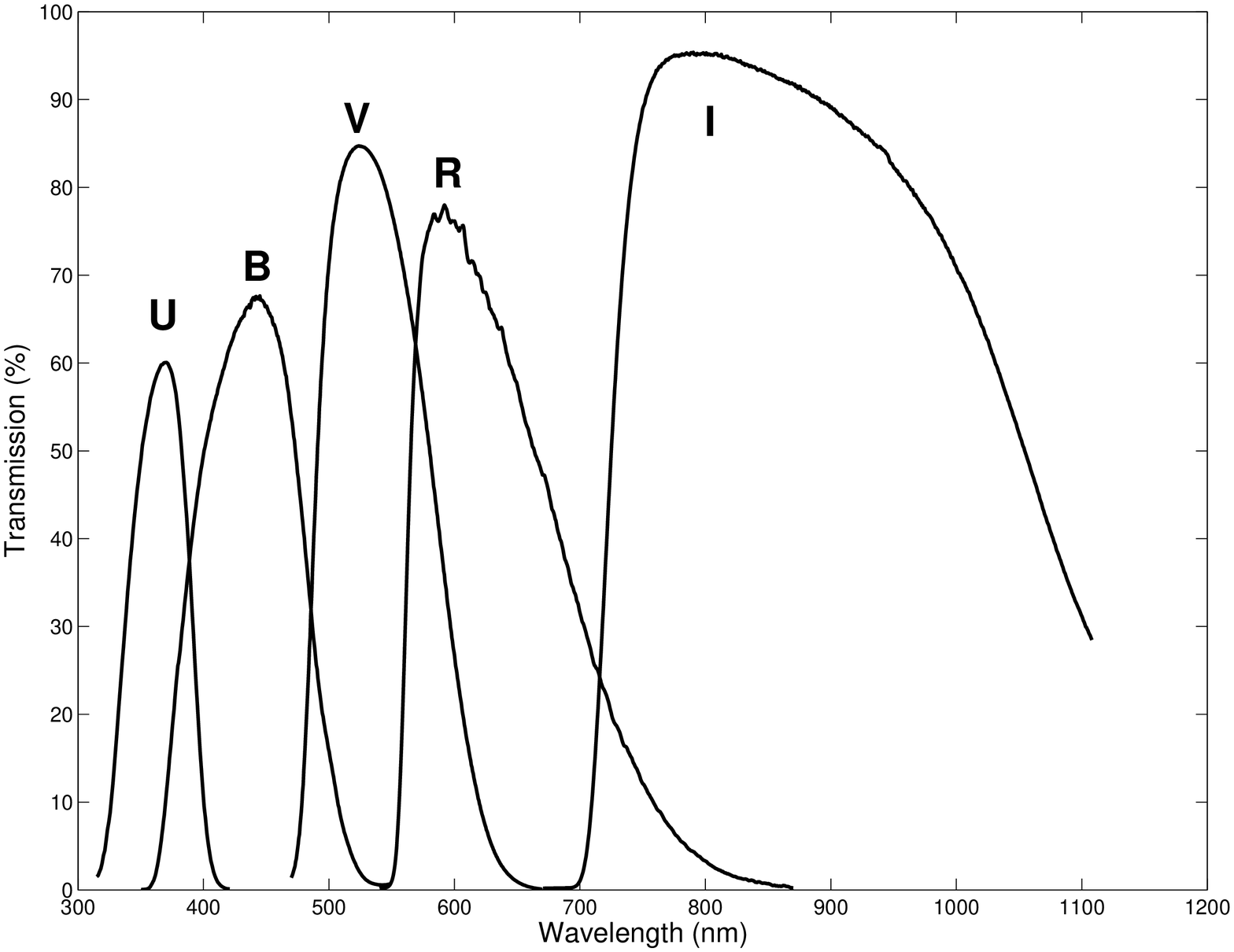}{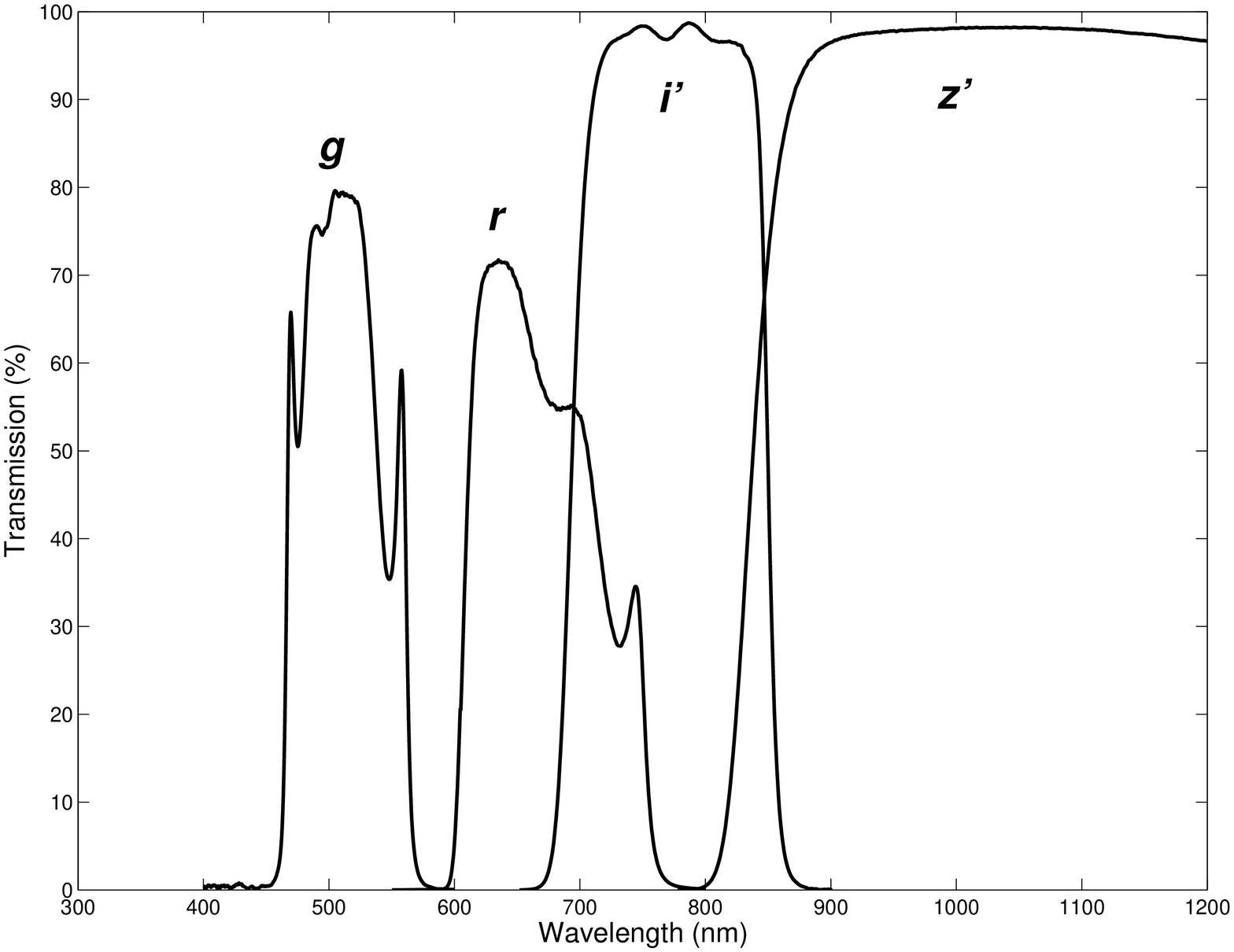}
     \caption[P60 Broadband Filter Transmission Curves]
     {P60 Broadband Filter Transmission Curves.  The top plot shows the
     Johnson $U$, $B$, $V$ and Kron $R$, $I$, while the bottom plot
     shows Gunn $g$ and $r$ and Sloan $i^{\prime}$ and $z^{\prime}$.
     These results can be found in tabular form online at
     http://www.astro.caltech.edu/\~{}ams/P60/filters.html.}
\label{fig:p60filters}
\end{figure}

Our typical filter wheel configuration consists of a set of standard broadband 
filters: Johnson $U\,B\,V$ (\citealt{b90} and references therein),
Kron $R\,I$ (functionally similar to Cousins $R_{\mathrm{C}}I_{\mathrm{C}}$,
\citealt{b90}), Sloan $i^{\prime}\,z^{\prime}$ 
\citep{fig+96}, and Gunn $g$ \citep{tg76}; 
two variations on Sloan $z^{\prime}$: 
$z_{\mathrm{short}}$ and $z_{\mathrm{long}}$;
and two narrow band H$_{\alpha}$
filters ($\lambda_c / \Delta \lambda =$ 6564/100, and
6584.65/17.5).  We have found significant deviations from
the canonical transmission curves for some of our broadband filters.  
We therefore measured the 
transmission curves of all of our broadband filters, and the results are shown
in Figure \ref{fig:p60filters}.  
These measurements are also available in tabular
form online at http://www.astro.caltech.edu/\~{}ams/P60/filters.html. 

\subsection{Observatory Conditions}
\label{sec:conditions}
Observing conditions at Palomar are highly seasonally dependent.  
In the summer months,
it is rare to lose an entire night due to weather.  The average
seeing at the P60 in the summer is $\sim 1.1$\arcsec\ in $R$-band.  The winter
months, however, are much worse.  As an extreme example, the P60 was closed for
15 full nights in January 2005.  Average seeing degrades to $\sim 1.6$\arcsec,
and can at times be significantly worse.  The seeing we experience at the P60 
is often times slightly worse (by $\sim 0.2$\arcsec) 
than the values reported at the 
200'' Hale Telescope.  We attribute this 
primarily to the difficultly
we have encountered determining and maintaining an accurate focus value (see
\S\ref{sec:efficiency}).

Sky backgrounds levels are generally good at Palomar, although they
have increased somewhat over the last decade as the area has become more
populated.  In recent images at P60 with the new CCD we have found sky 
background levels of 19.9, 19.0, 18.8, and 17.7 mag per PSF (here
approximated as a circular aperture of 1.5\arcsec diameter) 
in $B$, $V$, $R$, and $I$
respectively. 
The three-sigma limiting magnitudes of our current system are 
20.5 mag in $B$, $V$, and $R$, and 19.8 mag in $I$-band for an isolated point
source in a one minute exposure.
These results are summarized in Table \ref{tab:performance}.

The shortest recommended exposure time is set by the shutter mechanism.  
For exposures shorter than two seconds, the shutter speed becomes 
important and the true opening time (measured from a flat-field linearity
curve) is not strictly repeatable.  The longest recommend exposure is 
limited by the fact that we are not using a guider to assist in telescope
tracking.  This value is therefore dependent upon external conditions.  
In standard
seeing of 1.5\arcsec,
exposure lengths longer than 180 seconds begin to show image
degradation.  Under good seeing conditions of 1.0\arcsec, we have noticed
degradation in images longer than 90 seconds.  Users requiring deep images
of a field will need to split up their observations into exposures of this
length, and thereby sacrifice readout overhead.

\subsection{Observatory Efficiency}
\label{sec:efficiency}
The P60 currently devotes on average $\approx$ 50\% of the time the dome
is open for observations to science exposures.  This value is quite
variable, however, depending primarily on the number of different fields
observed each night.
An overview of the typical nightly efficiency is presented in Table 
\ref{tab:efficiency}.  Please note the values presented are given in terms
of the total time the dome is open, not the total available dark time.
Additional factors such as weather can affect the overall efficiency 
significantly.

Besides required operations such as telescope slews, the primary constraint
on our system efficiency comes from focusing.
We have found the secondary mirror on the telescope
to be unstable, particularly at higher elevations.  Large
telescope slews unpredictably alter the secondary mirror
position, thereby taking the telescope out of focus.  
While engineering work to reinforce the structural support of the secondary
in the Spring of 2006 has improved stability, we still conduct  
a focus loop every time we slew to a new target to maintain focus (this loop
is disabled for rapid-response observations).  As each individual focus
loop takes $\approx$ 3 minutes, visiting a large number of fields each night
can have a significant impact on our system efficiency. 

\begin{deluxetable}{lrrrr}
  \tabletypesize{\footnotesize}
  \tablecaption{P60 On-Sky Performance}
  \tablecolumns{5}
  \tablewidth{0pc}
  \tablehead{ & \colhead{$B$} & \colhead{$V$} & \colhead{$R$} & \colhead{$I$}
            }
  \startdata
        Sky Brightness & \nodata & \nodata & \nodata & \nodata \\
        ~~~~~mag per arcsec$^{2}$ & 20.8 & 19.9 & 19.6 & 18.6 \\
        ~~~~~mag per arcmin$^{2}$ & 11.9 & 11.0 & 10.8 & 9.8 \\
        ~~~~~mag per PSF\tablenotemark{a} & 20.1 & 19.2 & 19.0 & 18.0 \\
        Limiting Magnitude & 20.5 & 20.5 & 20.5 & 19.8 \\
  \enddata
  \tablenotetext{a}{We approximate our PSF here as a circular aperture of
        diameter 1.5\arcsec.}
\label{tab:performance}
\end{deluxetable}

\begin{deluxetable}{lr}
  \tabletypesize{\footnotesize}
  \tablecaption{P60 Nightly Efficiency}
  \tablecolumns{2}
  \tablewidth{0pc}
  \tablehead{\colhead{Property} & \colhead{Time Spent}
            }
  \startdata
        Science Exposures & 53\% \\
        Focusing & 12\% \\
        Readout Time & 8\% \\
        Photometric Standards & 4\% \\
        Scheduler Calculations & $<$ 1\% \\
        Other\tablenotemark{a} & 23\% \\
        \hline Total & 100\% \\
  \enddata
  \tablenotetext{a}{``Other'' includes all additional system components, such
        as telescope motion, changing filters, adjusting focus, and
        gathering status information.  Because most of these operations
        are done in parallel, it is impossible to disentangle each
        individual contribution.}
\label{tab:efficiency}
\end{deluxetable}

Additionally, 
our relative efficiency is lowered by $\lesssim $ 5\% because the P60 is
not equipped with a guider.  As mentioned in \S  
\ref{sec:conditions}, this puts an upper limit on suggested exposure
times.  In many cases we must use shorter exposures than would
otherwise be optimal to minimize the fraction of time spent in CCD
readout.  We note, however, that real-time scheduling has no 
noticeable impact on efficiency, as the OSS spends less than 1\% of the 
available time each night calculating which target to observe next. 

\subsection{Transient Response Time} 
\label{sec:response} 

The telescope response time to transient notices currently varies from
2--6 minutes.  Our fastest response time was for GRB\,050906, for which
we began observations 101 s after receiving the trigger notice (114 s
after the GRB; \citealt{GCN.3931}).  Under the current system, observations
of transient events do not begin until the previous observation has 
successfully completed.  Even though most exposures are relatively short,
this could take up to 5 minutes and explains why we have not met our 
stated response time goal in all cases.  We are currently in the process
of implementing an instantaneous interrupt capability, and aim to improve the
response time to $< 3$ minutes by the end of Summer 2006.

\section{Conclusions}
\label{sec:conclusions}

In this paper we have presented our efforts to automate and roboticize
the Palomar 60-inch telescope.  As of September 2004, all components of the
system operate in a fully automated fashion, making P60 one of the few
robotic, medium-aperture facilities in the world.  The P60 has been
routinely responding to \Swift\ GRB alters over the last year and a half,
and will continue to do so over the lifetime of the \Swift\ mission.  
The system is well-positioned for the plethora of optical transients
that will be discovered in the upcoming years.

In addition to the current optical camera, we are planning several 
major upgrades to further improve the scientific capabilities of the system.
In the near-term, our top priority is to add a near-infrared (NIR) camera
to P60.  We have already acquired the NIR detector from the out-of-use
Cerro Tololo Infrared Imager (CIRIM\footnote{http://www.ctio.noao.edu/instruments/ir\_instruments/cirim/cirim.html}) and upgraded the controller electronics.
We are currently working on both optical design and software development,
with the hope of having both cameras mounted and functional in the next year.
We also plan to make the P60 fully compliant with the Virtual Observatory
Event Network (VOEventNet\footnote{http://voevent.net}) protocol.  
In this manner the system can communicate 
with other observatories around the world without any needed human 
intervention.

As longer term projects, we are exploring the possibility of adding either
a polarimeter or a multi-band camera to the facility.  Regardless of the 
details, we are committed to making the P60 a scientifically productive
facility in the years to come.

\acknowledgments{We would like to thank the entire staff at Palomar 
Observatory, without whose patience and hard work this project would not
have been possible.  S.~B.~C.~and A.~M.~S.~ are supported by the NASA
graduate Student Research Program.  A.G.~acknowledges support by NASA through
Hubble Fellowship grant HST-HF-01158.01 awarded by STScI.  GRB research
at Caltech is supported through NASA and the NSF.}


\end{document}